\documentstyle[aps,pre]{revtex}

\newcommand{\be}{\begin{equation}}
\newcommand{\ee}{\end{equation}}
\newcommand{\BE}{\begin{eqnarray}}
\newcommand{\EE}{\end{eqnarray}}
\newcommand{\vp}{\varphi}
\newcommand{\e}{\epsilon}

\newcommand{\barr}{\begin{array}} 
\newcommand{\earr}{\end{array}}

\input psfig.sty
\begin{document}

\title{Helicoidal model for DNA opening.}
\author{Maria Barbi$^{1,3}$, Simona Cocco$^{2,3}$ and Michel Peyrard$^3$}

\address{$^1$ Istituto Nazionale di Fisica Nucleare, Dipartimento di Fisica, Universit\'a degli Studi di
Firenze, Largo E. Fermi, 2 - 50125 Firenze, Italy}

\address {$^2$  Dipartimento di  Scienze Biochimiche
Universit\'a di Roma "La Sapienza"
P.le A. Moro, 5 - 00185 Roma, Italy}

\address{$^3$ Laboratoire de Physique, CNRS URA 1325, Ecole Normale
Sup\'erieure de Lyon, 46 All\'ee d'Italie, 69364 Lyon Cedex 07,
France.}
\date{\today}

\maketitle

\begin{abstract}
We present a new dynamical model of DNA. This model has two degrees of
freedom per base-pair: one radial variable related to the opening of
the hydrogen bonds and an angular one related to the twisting of each
base-pair responsible for the helicoidal structure of the molecule.
The small amplitude dynamics of the model is studied analytically :
we derive small amplitude envelope solutions made of 
 a breather in the radial variables combined with a kink in
the angular variables, showing the role of the topological constraints
associated to the helicoidal geometry.
We check the stability of the solutions  by numerical integration
 of the  motion equations. 

\end{abstract}


\section{Introduction.}

Simple dynamical DNA models have been recently studied \cite{Gaeta} to
understand the possible basic mechanisms of some biological processes
such as denaturation, initiation of the transcription, and
transcription.  The thermal denaturation has been investigated by
Peyrard, Bishop and co-workers \cite{Peyrard1,Peyrard5} with a
unidimensional model (PB model) which allows local openings of the
hydrogen bonds and formation of denaturation bubbles.  The local
openings can be analytically described as breather-like objects of
small amplitude, which have nevertheless interesting properties: 
as long as their amplitude is small enough, they can move along the
chain, collect energy and grow \cite{Peyrard7}.  They can also be
trapped by some local dishomogeneities \cite{Peyrard6}, which suggests
that the properties of breathers could allow the formation of the
transcription bubble after the interaction with the bound
RNA-polymerase.

However, in a realistic description of DNA dynamics, the topological 
constraints related to the helicoidal structure of the molecule cannot 
be ignored. 
Activation or repression effects  caused by conformational changes 
and, in particular, prevention of transcription due to a large positive
excess in twist, are known and largely investigated in biology.
In fact,  during all processes in which the DNA base-pairs open, a 
local unwinding of the helix follows for topological reasons. 
Consequently a local extra-twist accumulates at the two ends of the
bubble and induces a long range elastic stress. 
Although the mechanical properties of DNA
have recently been the object of a renewed
interest\cite{Marko1,Marko2}, previous attempts to take into account
the helicoidal structure in the models have been limited to the
introduction of the forces that can appear due to the proximity in
space of bases which are  non-adjacent in the sequence \cite{Gaeta}.

Our aim here is to build a simple model that  takes into account
the twist-opening interactions due to the helicoidal molecular
geometry.  Such a model  provides  an extension
of the PB approach towards a more realistic
description of biological
processes. It can also be useful
  for more general studies of the interaction
between geometrical conformation and dynamical properties of the
molecule  referring to the recent mechanical experiments on
DNA \cite{SmithCui,StrickAllemand,ens}.

In designing such a model  each base will be considered as a
single, non-deformable object. 
We introduce  two degrees of freedom per
base-pair: one radial variable related to the motion of the bases
along the diameter that joins their attachment points to the
helicoidal backbone, and a twist angle of each base-pair defined by
the angle between  this diameter and a reference direction. This
twist angle, which increases from one base-pair to the next one, 
is responsible for the helicoidal structure of the molecule.  
From the resulting Lagrangian the equation of motion for
small displacements with respect to the equilibrium position are written.
We then derive 
analytical approximated solutions for the
small amplitude nonlinear distortions of the molecule.
 As suggested by the
helicoidal geometry we show that the radial
breather-like opening is associated to an angular un-twisting of the
molecule. 
We finally present the results of  numerical simulations of the dynamics 
showing that these solutions 
are stable for long periods of time and can move along the chain.

The analytical method of derivation of the envelope soliton solution
of the equation of motion used in this work has been proposed and developed 
extensively  in a previous paper \cite{noi}. 
The reader is referred to this work to better follow the technical steps of
 our calculation.


\section{The helicoidal model.}

In the PB model \cite{Peyrard1},
 the
bases are point masses
 allowed to move only in the direction of the hydrogen
bonds that connect them. A Morse potential describes the effect of
these bonds, while neighboring bases along the same strand are
harmonically coupled, simulating the stacking interactions. 

Similarly, in our model the group made of a sugar
ring and its connected base is treated simply as a point mass
(without distinction between the different base types); the
phosphate backbone between two base-pairs 
is modelised  as an elastic rod. The additional
 twist motion is now introduced by  allowing the two bases in
each pair to move in the base-pair plane instead of constraining them
on a line. It is convenient to choose a polar coordinate
system (Fig. 1).  The model does not attempt to describe the acoustic
motions of the molecule since only the stretching of the base-pair
distance is considered. This amounts to fixing the center of mass of
the base pair, i.e. the two bases in a pair are constrained to move
symmetrically with respect to the axis of the molecule.  Then, to describe
the stretching of a base pair and the variation of the helicoidal twist
we  need only two  degrees of freedom per base pair: the
coordinates ${r}_{n}$ and ${\varphi}_{n}$ of one of the two bases with
respect to a fixed reference frame. 

As in the PB model,  a Morse potential describes the hydrogen
bonds linking bases in a pair with an equilibrium distance $R_0$.

A proper choice of the coupling between radial and angular variables
has to reproduce the equilibrium helicoidal structure.  In 
DNA, the latter originates  from the competition between the
hydrophobic effect (that tends to eliminate water from the core of the
molecule by bringing the neighboring base-pair planes closer) the
electrostatic repulsion between neighboring base planes (which has the
opposite effect) and the rigidity of the two strands (that separates
the external ends of the base pairs by essentially a fixed length
related to the phosphate length) \cite{Calladine}.  We combine the
first two forces in a unique stacking effect that fixes base-plane
distance $h$ in the model.  As the
equilibrium backbone length $L$ is greater than $h$, it is then
necessary to incline the strands in the typical helicoidal structure to
minimize the energy. 

The geometrical parameters $h$ and $R_0$ have been chosen to match the
structure of B-DNA. Then, in order to impose an equilibrium twist
angle $ {\varphi}_{n} -{\varphi}_{n-1} =\pm {\Theta}_{0}$ 
between two consecutive base pairs, we select the
equilibrium length $L$ of the springs representing the phosphate
backbone to be

\begin{eqnarray}
\label{L}
L
= \sqrt{{h}^{2}+ 4{R}_{0}^{2}\sin^2\big({{\Theta}_{0}\over 2}\big) }\, >\, h
\end{eqnarray}
\smallskip

However, because of the invariance of $L$ with respect to the sign of
${\Theta}_{0}$, the elastic rods rigidity 
is not sufficient to guarantee the correct
helicoidal shape. A {\it zig-zag} structure with
a random succession of  $\pm {\Theta}_{0}$ 
 would as well minimize the energy of the system.
This can be avoided by adding a three-body curvature term to 
the Lagrangian, that imposes a continuity between the differences 
${\varphi}_{n} -{\varphi}_{n-1}$  and  ${\varphi}_{n+1} -{\varphi}_{n}$.

The final Lagrangian  is written considering the expression of the 
three-dimensional distance between the two bases along the strand 
(Fig. 1); it reads

\begin{eqnarray}
\label{HBC}
{\cal L} & =  
& \sum_{n} 
\big( m{\dot{{r}_{n}}}^{2}+m{{r}_{n}}^{2}{\dot{{\varphi}_{n}}}^{2}\big) 
- D \big( e^{-\alpha({r}_{n} - {R}_{0})} -1 \big)^2  
\nonumber \\
&-& \sum_{n} K {\big( \sqrt{{h}^2 + {r}_{n-1}^2 + {r}_{n}^2 
- 2 {r}_{n-1} {r}_{n} \cos ({\varphi}_{n} - {\varphi}_{n-1})} - L \big)}^2
  \nonumber \\
&-& \sum_{n}
G_0\,{\big( {\varphi}_{n+1}+{\varphi}_{n-1}-2{\varphi}_{n} \big)}^{2}
\; 
\end{eqnarray}
where 
$m$ is the base mass, $D$ and $\alpha$ are the depth and width of the Morse
potential well (without distinction between double and triple hydrogen
bonds), $K$ is  the backbone elastic constant   and $G_0$  the backbone
curvature constant. 

The results presented below have been obtained with, 
   $m=300 u.m.a.$, $D=0.04 eV$, $\alpha=4.45$ \AA$^{-1}$
that are the same valued adopted in the PB
model,
and with $K=1.0 eV $ \AA$^{-2}$  and $G_0=K R_0^2 / 2$.
We intend to refine these values  in future 
statistical mechanics studies by comparing the predictions
of the present model with some available experimental data,
{\em eg.} the temperature of DNA denaturation and the rigidity
of the molecule \cite{Mezard}. 

For the geometrical parameters we adopt the B-DNA 
values $R_0 \approx 10.0$ \AA, ${\Theta}_{0}=36^{\circ} $ and $h=3.4$ \AA.


One can simplify the Lagrangian by introducing the
adimensional variables $r'_n=\alpha r_n$,  the rescaled time 
$t'=\sqrt{D \alpha^2/m}\, t$ and the renormalized parameters 

\begin{eqnarray}
    K'&=&K/D \alpha^2  \nonumber \\ 
    {\cal G'}&=&G_0/D \nonumber \\
    R'_0&=&\alpha R_0  \nonumber \\
    h'&=&\alpha h      \nonumber \\
    L' &=& \alpha L\,.
\end{eqnarray}

In the following the new variables will be written without primes.    

We perform an expansion of the energy of the backbone springs in
(\ref{HBC}) up to the second order around
 the equilibrium position:

\BE
y_n        &=&    r_n-R_0\,\\
{\phi}_{n} &=&R_0 (\,{\vp}_{n}-n{\Theta}_{0}\,)\,,
\EE
\smallskip
 
obtaining

\begin{eqnarray}
\label{appspring}
{\big( \sqrt{{h}^2 + {r}_{n-1}^2 + {r}_{n}^2 - 
2 {r}_{n-1} {r}_{n} \cos ({\varphi}_{n} - {\varphi}_{n-1})} -L\big)^2} \simeq
\nonumber \\ 
 {{ R_0^2 }\over{L^2}}
{\big[  \sin{{\Theta}_{0}} ({\phi}_{n}-{\phi}_{n-1})
  +(y_n+y_{n-1})(1-\cos{\Theta_0}) \big]}^{2}  \,.
\end{eqnarray}
\smallskip

We also expand the Morse potential up to the fourth order
in  $y_n$ 

\begin{equation}
\big( e^{- {y}_{n}} -1 \big)^2 =
\frac{1}{2} y_n^2- \frac{1}{2} y_n^3 +\frac{7}{24} y_n^4 + O({y_n}^5)\,. 
\end{equation}

The difference in the order of the two expansions is consistent with
our parameter choice.

With these expansions, the equations of motion become:

\BE
\label{eqm1}
{\ddot y}_n 
&=&\big(1+{{y_n}\over{R_0}}\big){{1}\over{R_0}}{\dot{\phi}}_{n}^{2}
- (y_n- {{3}\over{2}} y_n^2 +{{7}\over{6}}y_n^3 ) 
-K_{yy} ({y}_{n+1} + {y}_{n-1} + 2 {y}_{n})      \nonumber \\
&&-{{K_{y\phi}}\over{2}}({\phi}_{n+1} -{\phi}_{n-1}) \\
\label{eqm2}
{\ddot{\phi}}_{n} 
&=&  K_{\phi \phi} ({\phi}_{n+1}+{\phi}_{n-1}-2{\phi}_{n}) 
\nonumber \\
&-&  G  ( {\phi}_{n+2}+{\phi}_{n-2}-4{\phi}_{n+1}-4{\phi}_{n-1}+6{\phi}_{n})
\nonumber \\
&&+{{K_{y\phi}}\over{2}}({y}_{n+1} -{y}_{n-1})- 
\frac{2}{R_0}{\dot{y}}_{n}{\dot{\phi}}_{n}-
\frac{2}{R_0}{{y}}_{n}{\ddot{\phi}}_{n}-
\frac{2}{R_0^2}y_n {\dot{y}}_{n}{\dot{\phi}}_{n}
-\frac{1}{R_0^2}y_n^2 {\ddot{\phi}}_{n}
\EE

where

\BE
K_{yy} &=&  \big( K R_0^2 / L^2 \big){(1 - \cos{\Theta}_{0})}^{2}\;,\\
K_{\phi\phi} &=&  \big( K R_0^2 / L^2 \big){(\sin^2{\Theta}_{0})}\;, \\
K_{y\phi}&=&2 \big(K R_0^2 / L^2 \big)(\sin{\Theta}_{0})(1-\cos{\Theta}_{0})\; \\
G &=& {\cal G} / R_0^2 \, .
\EE

The constants $K_{yy},K_{\phi \phi}$ are the effective elastic constants respectively
for the base pairs opening and the twist rotation, $K_{y\phi}$ is the 
coupling constant between stretching and twist; 
these three constant are geometrically related.


\section{Localized breather-like solutions.}

We now look for nonlinear localized solutions, characterized by the
propagation of a coherent collective structure on a time scale
greater than the time scale of the vibrations of each particle around
its equilibrium position.
 
According to the method developped in \cite{noi}  we first
solve for a  wave packet solution of the linearized  system
 with weak dispersion. 
This amounts to solving
\be
\label{d}
(\hat{J}(q)- \omega^{2}_{l}(q))\vec{V}_{l}(q) =0 
\ee 
 where  $l=+,-$ is  the branch index and
\be 
\hat{J}(q)= 
\left(\matrix{1+2K_{yy}(1+\cos{q}) & iK_{y  \phi} \sin{q}  \cr 
-iK_{y  \phi} \sin{q} & 2 K_{\phi \phi}(1-cos{q}) + G( 6-8\cos{q}+2\cos{2q})}\right)
\equiv \left(\matrix{a & c  \cr 
c^* & b}\right) \label{reldisp}
\ee
finding  the 
eigenvalues and eigenvectors
\be 
 \omega^2_{\pm}(q)=\frac{1}{2}(a+b\pm\sqrt{(a-b)^2+4|c|^2}\;)
\ee

\be  
\label{autovettori1}
\vec{V}^{\pm}(q)= {\cal N_{\pm}} \left( \barr{c} 1 \\  \frac{a-\omega_{\pm}^2}
{-c}
  \earr \right)
\ee
where ${\cal N_{\pm} }$  are the vector norms.

Then we apply a perturbative expansion of the system (\ref{d})
around one normal mode. We have chosen as carrier wave a mode
($q_0,\omega_{+}=\omega_{+}(q_0)$) on the
optical branch and with a small wave number
that corresponds to a
prevalently radial excitation weakly oscillating.
 In fact, keeping in mind DNA opening, we shall calculate   the
behaviour of a localized bubble-like radial distorsion and the induced
angular dynamics.  
We obtain the wave-packet  group velocity from the first
order eigenvalue correction
\be 
\omega_{+}^{(1)}=   \frac{ {\vec{V}^{+*}} \hat{J'} \vec{V}^{+}}{2 \omega_{+}}=
\frac{1}{2 \omega_{+}}(a' |{V}^{+}_1|^2+c'{V}^{+*}_1{V}^{+}_2+
 c'^{*}{V}^{+*}_2{V}^{+}_1+b'|{V}^{+}_2|^2) \quad ,
\ee
where the primes indicate the derivatives with respect to $q$ calculated
at $q_0$. 
We then obtain  the correction to the eigenvector $\vec{V}^{+}$
\be
\vec{V}^{(1)}=\alpha \vec{V}^{-} 
\ee
\be
\alpha=  \frac{ {\vec{V}^{-*}} \hat{J}' \vec{V}^{+}}
{\omega_{+}^2-\omega_{-}^2}=
\frac{a' {V}^{-*}_1 {V}^{+}_1 +c'{V}^{-*}_1{V}^{+}_2+
 c'^*{V}^{-*}_2{V}^{+}_1+b'{V}^{-*}_2 {V}^{+}_2 }{\omega_{+}^2-\omega_{-}^2}
\ee
and  from the second order eigenvalue correction
  the group velocity dispersion is obtained as

\BE
\label{os}
\omega_{+}^{(2)}&=& \frac{1}{\omega_{+}}
 ({\vec{V}}^{+*} \frac{\hat{J}''}{2} \vec{V}^{+}
-{\omega_{+}^{(1)}}^2+ 
  \frac{| {\vec{V}^{-*}} \hat{J}' \vec{V}^{+}|^2} { \omega^2_{+}
-\omega^2_{-}})\ \nonumber \\
&=& \frac {1}{\omega_{+}}(\frac{1}{2}(a'' |{V}^{+}_1|^2+c''{V}^{+*}_1{V}^{+}_2+
 c''^*{V}^{+*}_2{V}^{+}_1+b''|{V}^{+}_2|^2)-
{\omega_{+}^{(1)}}^2 +|\alpha|^2 (\omega_{+}^2-\omega_{-}^2))
\EE

We now take into account the nonlinearity. We look for a small amplitude
solution. The expansion parameter $\epsilon$ for the
solution is introduced to solve the equation of motion at increasing order of
accuracy inserting the nonlinear terms in a progressive way.
This iterative expansion is combined with the expansion in multiple
scales for the  weak dispersive wave packet. The latter requires the
introduction of the variables $x_1=\epsilon x,
t_1= \epsilon t, t_2=\epsilon^2 t$ for the slowly varying amplitudes,
where $\epsilon$  is the same parameter as before.
We look then for a solution of the form
\BE
\label{wpc}
\vec{E}(n)&=&\epsilon e^{i(q_0n_0-\omega_{+}t_0)}  
 (\vec{V}^{+}-i \epsilon \vec {V}^{(1)} \frac{\partial }{\partial x_1})
 A(x_1,t_1,t_2)+\epsilon \vec\sigma(x_1,t_1,t_2) \nonumber \\
&+&\epsilon^2
 e^{2i(q_0n_0-\omega_{+}t_0)} 
\vec\gamma(x_1,t_1,t_2)
+\epsilon^2 \vec\mu(x_1,t_1,t_2) \,
\EE
where the first term is the wave packet,
 the second term
 arise because the linear system  admits a 
 constant solution with a non zero 
second component, corresponding to the null column of $\hat{J}(0)$.
The constant and second harmonic $\epsilon^{2}$ terms are
induced by the quadratic nonlinearity in the equation of motion.

We have now to  determine the slowly variyng amplitudes   
$ A(x_1,t_1,t_2),\vec\sigma(x_1,t_1,t_2),
\vec\gamma(x_1,t_1,t_2),\vec\mu(x_1,t_1,t_2)$ 

From the $O(\epsilon^2)$  equation of motion we obtain for
the second harmonic terms the system of equations:

\be
(\hat{J}(2q_0)- 4\omega_{+}^2)\vec{\gamma}=
\left(\matrix{3/2 {V_1^{+}}^2 - \frac{\omega_{+}^2}{R_0}{V_2^{+}}^2   \cr 
\frac{4\omega_{+}^2}{R_0}V_1^{+} V_2^{+}} \right) \;A^2 \; .
\ee
Its solution may be written as $\vec{\gamma}=\vec{\gamma_c} A^2$
and for the constant terms
\be
\hat{J}(0)\vec\mu -i\hat{J'}(0)\frac{\partial}{\partial x_1} \vec\sigma
=\left(\matrix{3 |{V_1^{+}}|^2 + \frac{2\omega_{+}^2}{R_0}|{V_2^{+}}|^2   \cr
 0} \right)\;|A|^2 \quad .
\ee
Inserting the value of $\hat J (0)$ from (\ref{reldisp}) in the above
matricial equation, we obtain the identity
\be
\label{mu1}
(1+4K_{yy})\mu_1+K_{y\phi}\frac{\partial}{\partial x_1}\sigma_2=
(3 |{V_1^{+}}|^2 + \frac{2\omega_{+}^2}{R_0}|{V_2^{+}}|^2)  \;|A|^2
\ee
that does not allow to determine the two unknown variables.
We have thus to consider the $O(\epsilon^3)$ system of   equations 
for the constant terms
\be
-i\hat{J'}(0) \frac{\partial}{\partial x_1}\vec\mu-
\hat{J''}(0) \frac{\partial^2}{\partial x_1^2} \vec\sigma=
\left(\matrix{ \frac{2i\omega_{+}}{R_0}|{V_2^{+}}|^2(
A^*   \frac{\partial A}{\partial t_1}-\frac{\partial A^*}{\partial t_1}A)
\cr  \frac{-2i\omega_{+}}{R_0} 
({V_2^{+}}^* V_1^{+}-{V_1^{+}}^* V_2^{+})
\frac{\partial}{\partial t_1} |A|^2}\right)
\ee
corresponding to the two equations:
\BE
\label{mu2}
K_{y  \phi} \frac{\partial}{\partial x_1}\mu_2 
&=& \frac{2i\omega_{+}}{R_0}|{V_2^{+}}|^2(
A^*   \frac{\partial A}{\partial t_1}-\frac{\partial A^*}{\partial t_1}A) \\
\label{sigma2}
-K_{y\phi}\frac{\partial}{\partial x_1}\mu_1-
2K_{\phi \phi} \frac{\partial^2}{\partial x_1^2} \sigma_2 &=&
 \frac{-2i\omega_{+}}{R_0} 
({V_2^{+}}^* V_1^{+}-{V_1^{+}}^* V_2^{+})
\frac{\partial}{\partial t_1} |A|^2
\EE

The equation (\ref{mu2}) gives $\mu_2$ as a function of $A$.
The equation (\ref{sigma2}) can be integrated using
the wave packet property
$\frac{\partial A}{\partial t_1}=
- \omega_{+}^{(1)}\frac{\partial A}{\partial x_1}$. We
 then obtain from (\ref{mu1}), (\ref{sigma2})  the following system for
 $\mu_1$ and $\sigma_2$,
\be
\left(\matrix{ 1+4K_{yy} &K_{y  \phi} \cr -K_{y  \phi}
 &-2K_{\phi\phi}}\right)\;
\left(\matrix{\mu_1 \cr \frac{\partial \sigma_1}{\partial x_1}}\right)=
\left(\matrix{3 |{V_1^{+}}|^2 + \frac{2\omega_{+}^2}{R_0}|{V_2^{+}}|^2
\cr \frac{2i\omega_{+}\omega_{+}^{(1)} }{R_0} 
({V_2^{+}}^* V_1^{+}-{V_1^{+}}^* V_2^{+})}\right)\;|A|^2
\quad , \ee
the solutions of which may be written as 
\BE
\mu_1={\mu_1}_c |A|^2 \\
\sigma_2=\sigma_c\int|A|^2 dx_1
\EE

The nonlinear $O(\epsilon^3)$ terms in $\exp{i(q_0n_0-\omega_{+}t_0)}$ 
 in the equation of motion give rise to the nonlinearity that
balances the group velocity dispersion of the wave packet.
We indeed obtain the following 
Non Linear Schr\"odinger (NLS) equation for the envelope $A$
 expressed in a frame
moving at velocity $\omega_{+}^{(1)}$ 
(variables $S = x_1-\omega_{+}^{(1)} t_1,\tau=t_2 $):

\be
\label{NLScr}
i A_{\tau} + P A_{SS} + Q {|A|}^2 A = 0 
\ee
with
\BE
P &=& \frac{\omega_{+}^{(2)}}{2} \;, \\
Q &=& {{{ (Q_1 {V_{1}^+}^* + Q_2 {V_{2}^+}^*)  }} 
\over
{ 2  \omega_+ } }\;,
\EE
where
\BE
Q_1 &=& -\frac{7}{2}|V_{1}^+|^2 V_{1}^+ +\frac{\omega_{+}^{2}}{R_0^2}
(2 V_{1}^+ |V_{2}^+|^2 -{V_{1}^+}^* {V_{2}^+}^2)+\nonumber\\
&&3{V_{1}^+}^* {\gamma_1}_c+\frac{4\omega_{+}^{2}}{R_0}{V_{2}^+}^*{\gamma_2}_c
+3 V_{1}^+{\mu_1}_c \\
Q_2 &=& \frac{\omega_{+}^{2}}{R_0}(4{V_{1}^+}^* {\gamma_2}_c-
2{V_{2}^+}^* {\gamma_1}_c)+
 \frac{2\omega_{+}^{2}}{R_0} V_{2}^+{\mu_1}_c - \nonumber\\
&& \frac{2\omega_{+}^{2}}{R_0^2} {V_{1}^+}^2 {V_{2}^+}^* +
\frac{\omega_{+}^{2}}{R_0^2}(2|V_{1}^+|^2 V_{2}^+ +{V_{1}^+}^2 {V_{2}^+}^*)
\quad . \EE

If $PQ > 0$, equation (\ref{NLScr}) has an envelope soliton solution
\be
\label{F}
A(S,\tau) = {\cal A}\ \mbox{sech}{ [  {{1}\over{L_e}} (S-u_e \tau)]} 
                      \exp{ [ i {{u_e}\over{2P}} (S-u_c \tau)]}
\ee
where
\BE
{\cal A} = \sqrt{ {{u_e^2 - 2 u_e u_c}\over{2PQ}}  } \\
L_e = {{2P}\over{\sqrt{u_e^2 - 2 u_e u_c}}} 
\EE
are respectively the amplitude and the width of the curve.

Once the NLS equation is solved for $A(S, \tau )$,
we have $\sigma_2$,  $\vec{\gamma}$,
$\vec\mu$ (solving (\ref{mu2}) we obtain 
$\mu_2={\mu_2}_c \int  |A|^2 dx_1$) and then 
the complete solution 

\BE
y &=&   \e (V_1^+ -i\epsilon V_1^{(1)} \frac{\partial}
{\partial x_1}) A e^{i \theta} + c.c.
+\e^2 {\gamma_1}_c A^2 e^{2 i \theta} + c.c.
+\e^2 {\mu_1}_c {|A|}^2 +O(\e^3)
\\
\phi &=&  \e \sigma_c \int {|A|}^{2} dx_1 +
\epsilon ( V_2^+ -i\epsilon V_2^{(1)} \frac{\partial}
{\partial x_1}) A e^{i \theta} + c.c.
 +\e^2 {\gamma_2}_c A^2 e^{2 i \theta} + c.c.+
\e^2 {\mu_2}_c \int {|A|}^{2} dx_1  +O(\e^3)
\, .
\EE
 where $\theta= (q_0n_0-\omega_{+}t_0)\,.$
The  final result can be rewritten in the following form:
\BE
\label{y}
y &=& 
2\e {V_1}^+ {\cal A} \mbox{sech}{ [ \eta (x-V_e t)]} 
\cos{( {\cal K}x - {\Omega}t )} 
\nonumber \\ 
&+& \e^2 {V_1}^{(1)} {\cal A} \mbox{sech}{ [ \eta (x-V_e t)]} 
\left[\frac{-2}{ L_e}\tanh{(\eta(x - V_e t))} 
\sin{( {\cal K}x - {\Omega}t )} 
+{{u_e }\over{P}}\cos{( {\cal K}x - {\Omega}t )}\right]
\nonumber \\
&+& 2\e^2   {\gamma_1}_c {\cal A}^2 {\mbox{sech}}^2 { [ \eta (x-V_e t)]} 
\cos{(2 {\cal K}x - 2 {\Omega}t )}  
\nonumber \\ 
&+& {\mu_1}_c \e^2 {\cal A}^2 {\mbox{sech}}^2 { [ \eta (x-V_e t)]} +O(\e^3)
\EE
\BE
\label{phi}
\phi &=& 
\e (\sigma_c+\epsilon {\mu_2}_c) L_e {\cal A}^2 \tanh{(\eta(x - V_e t))}
-2\e {\cal A}|V_2^+| \mbox{sech}{ [ \eta (x-V_e t)]} 
\sin{( {\cal K}x - {\Omega}t )}
\nonumber \\ 
&-& \e^2 {V_2}^{(1)} {\cal A} \mbox{sech}{ [ \eta (x-V_e t)]} 
\left[\frac{2}{ L_e}\tanh{(\eta(x - V_e t))} 
\cos{( {\cal K}x - {\Omega}t )} 
+{{u_e }\over{P}}\sin{( {\cal K}x - {\Omega}t )}\right]
\nonumber \\ 
&-& 2 \e^2|{\gamma_2}_c| {\cal A}^2 {\mbox{sech}}^2 { [ \eta (x-V_e t)]} 
\big[ \sin{( 2{\cal K}x - 2{\Omega}t )} 
\big] +O(\e^3)
\EE
where
\BE
\eta &=& {{\e}\over{L_e}} \\
V_e &=& {\omega_+}^{(1)} + \e u_e \\
V_c &=&{\omega_+}^{(1)}  + \e u_c \\
{\cal K} &=& q_0 + {{\e u_e}\over{2P}} \\
{\Omega} &=& \omega_+ +  {{\e u_e}\over{2P}} V_c \, .
\EE

Fig. 3  shows the analytical solution (\ref{y}),(\ref{phi}) 
in the original variables (not renormalized), 
as a function of $x$ and $t$. Fig. 4 displays the 
results of numerical simulations of the system described by the
complete Lagrangian (\ref{L}) starting from the analytical
solution  (\ref{y}),(\ref{phi}) for $t=0$  as an initial
condition.
We have chosen 
 the carrier  wavevector $q_0=0.1$.

The shape of the solution is best seen on Fig. 3a, 3b which show 
for a few periods the  radial displacement $r_n(t)-R_0$ (Fig. 3a)
and of the  angular displacement $\varphi_n(t)$ (Fig. 3b), while
 Fig. 3c
 represents the twist angle, {\em i.e.} the difference
between neighboring  angles,
$\Delta\varphi_n(t)$. 
As in the PB model the radial motion has the shape of an asymmetric
breather,
the stretching of the base-pair distance being larger than its
compression, as expected from the asymmetry of the Morse potential.
The corresponding  angular motion clearly exhibits a kink
structure due to the non oscillating term $\sigma (x,t)$; a small
oscillating part due to the other terms in (\ref{phi}) is superimposed on it.

The shape of the solution agrees with the geometrical properties of
the molecule related to its helicoidal structure. It shows, as
expected, that our model can indeed describe the local untwisting which
should be coupled with the radial breather due to the geometrical
constraints.  To make this clearer we plot in Fig. 3c the evolution
of the twist angle. In addition to the local opening, small periodic
overtwists on the boundaries, coming from the oscillating part of the
solutions are visible.
The coupled breather and kink solutions move with a peak velocity 
$ V_e = \omega_{+}^{(1)} + \e u_e$  without changing their internal structure.

The long time evolution of the initial condition, calculated with the
full Lagrangian and shown in Fig.~4, attests that 
the local excitation  is very stable. It does not radiates energy
with the  exception of a small initial transient that could be
eliminated with absorbing boundary conditions.  
During all the simulation time 
  the total energy lost with the absorbing conditions is 
less than $ 10^{-5}$ of the total energy. It should be noticed 
that Fig. 4 does not show the actual oscillations due to the large
sampling time used to produce the graph. The slow oscillation visible
on Fig. 4 is due to a beating between the actual oscillation and the
sampling time. Fig. 4 shows in fact about 2230 
oscillations which proves the exceptional
stability of the breather in the helicoidal structure and the ability
of the geometrical nonlinearity and of the nonlinearity coming from
the Morse potential to keep the energy localized.

As our solution has been obtained with a small amplitude expansion, we
should not expect it to be a good solution when its amplitude
increases to much. This can be seen on the results of a numerical
simulation starting from a much larger amplitude (Fig.~5). In this
case the initial condition decays very fast and radiates energy before
stabilizing to a breather which has still a pretty high amplitude and
which is localized on about 6 sites (width at half height). 
Although it is very narrow the breather is still
mobile in the system and it is accompanied by a sharp kink as expected
from the geometrical constraints. The solution reached after the
transient is stable as shown in Fig. 5c: although the simulation has
been run with absorbing boundary conditions, the energy tends to
stabilize once the radiated waves have been dissipated at the ends of
the molecule.
In all the solutions described above the breather frequency $\Omega$ 
remains in the gap
between the two branches so that the breather does not emit acoustic
phonons.


\section{Conclusion}

In conclusion we have proposed a two-variable helicoidal DNA model
that, from a biological point of view, can be a better starting point
than the PB model for the studies of all the DNA dynamical processes
involving internal openings. Our model also describes the untwisting
that accompanies the base-pair stretching because of the   helicoidal
molecular geometry.
We analytically derive, using the method introduced in \cite{noi}, 
 a breather like solution
 in the 
radial variable  coupled with an angular
 solution dominated by a non oscillating 
   kink-like term.
A further step will be  the study of the formation of this kind of
excitation in presence of a thermal bath and of the 
onset of  thermal denaturation.
This will be useful in determining the model parameters by comparison
with denaturation temperature experimental data. Furthermore 
we expect to find interesting  thermodynamic properties 
due to long range effect generated by the helicoidal DNA structure.

As far as biological processes are concerned, the model may however still lack
some relevant features, in particular the possibility of bending and
stretching of the molecule. In order to obtain a stretchable molecule
axis we can simply replace the rigid constraint of fixed base planes
distance $h$ with an elastic term which leads to the introduction of a
third degree of freedom with a new elastic constant. This could be a
link to some recent experimental studies of the mechanical response
of a DNA molecule to some twisting and stretching forces  
\cite{SmithCui,StrickAllemand,ens}. 
Anyway the model contains in its  possible dynamical behaviors
some other important features, as {\em e.g.} the possibility of
local transitions from right hand B-DNA  to left hand Z-DNA,
related to the curvature term in the Lagrangian
depending on the constant $G_0$.
Our model is, finally,  a first attempt to insert in simple 
DNA dynamical model topological
constraints that are of great interest in today's biological studies.

\section*{Acknowledgments}

We would like to thank Prof. A. Colosimo and Prof. S. Ruffo for helpful
discussions. This work was initiated by discussions that took
place at the Institute for Scientific Interchange (ISI) of Torino, 
Italy, which is gratefully acknowledged.

\newpage

\begin{figure}[h]
\centerline{\psfig{figure=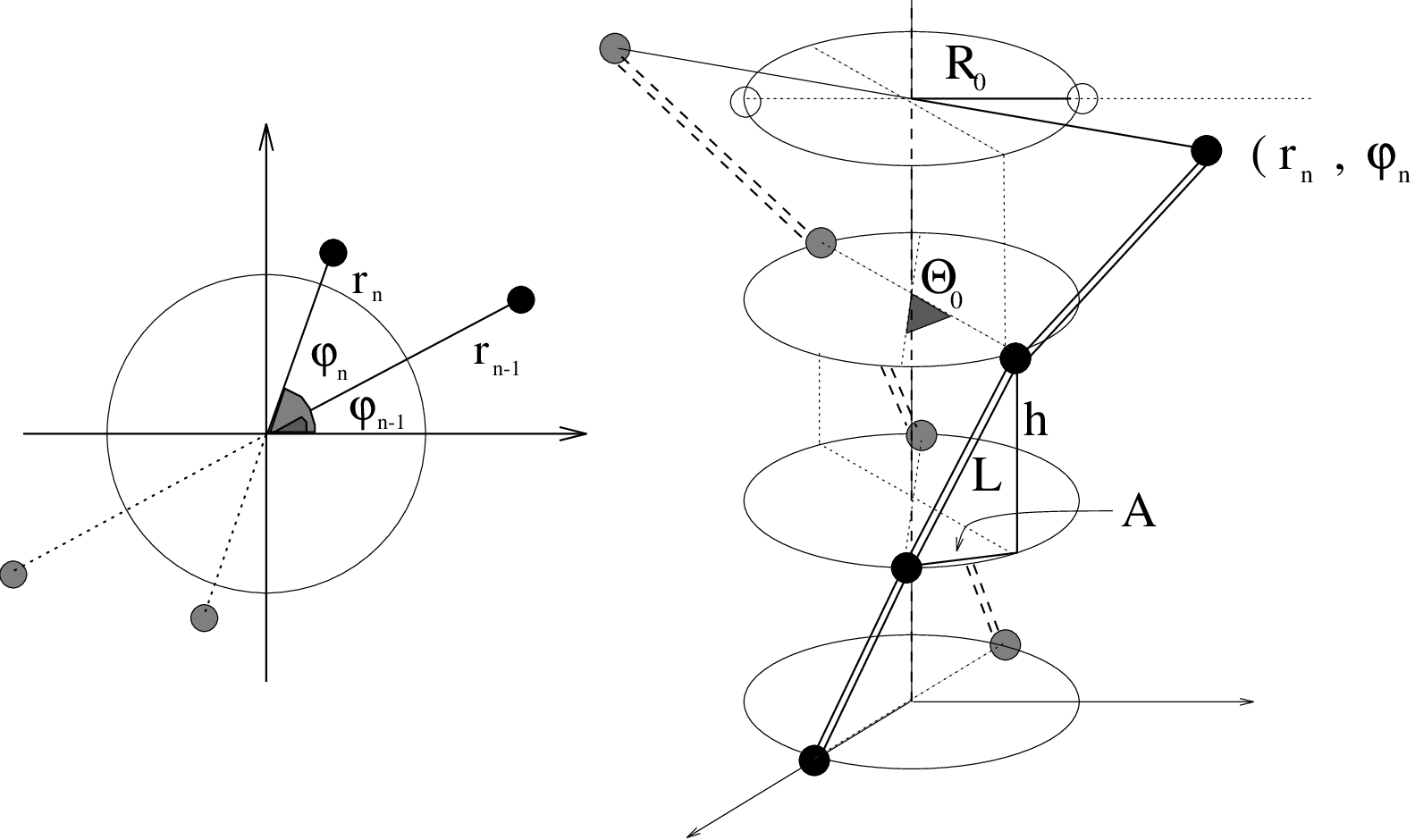,height=8cm}}
\vspace{0.5truecm}
\caption{Fig. 1: Choice of variables (left) and the schematic view
(right) of the model. The n-th base-pair is represented with its radial and
angular displacements $(r_n, \varphi_n)$ with respect to a fixed external 
reference frame. Neighbouring base-pairs are linked along the two strands
by elastic rods with equilibrium length $L$.
$R_0$, $\Theta_0$ and $h$ are the geometrical parameters of the model, 
with values deduced from the real B-DNA geometry.}
\end{figure}

\newpage

\begin{figure}[h]
\centerline{\psfig{figure=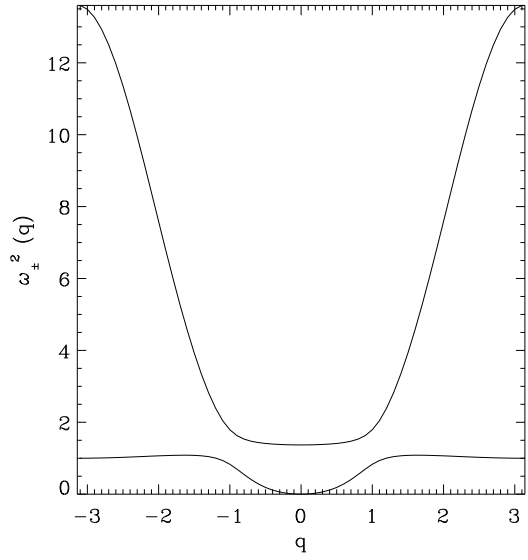,height=8cm}}
\vspace{0.5truecm}
\caption{Fig. 2: Dispersion curves for the linearized model with 
$K' = K/D\alpha^2 = 1.26$, ${\cal G'} = G_0/D = K'/2\,{R'}_{0}^{2}$.}
\end{figure}

\newpage

\begin{figure}[h]
\centerline{\psfig{figure=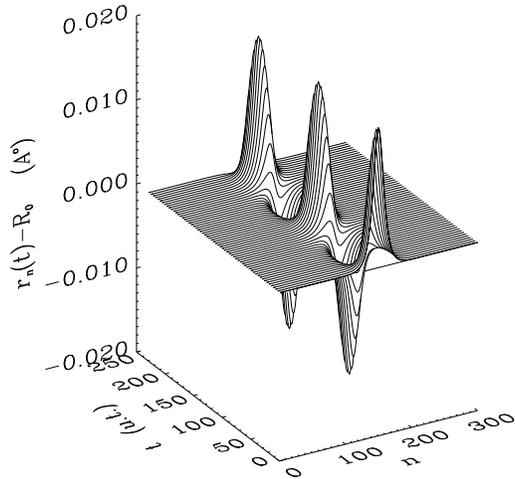,height=8cm}}
\vspace{0.5truecm}
\caption{Fig. 3a: Analytical solution (\ref{y}), (\ref{phi}): we show here the 
radial displacement $r_n(t)-R_0$, obtained with the model parameters 
indicated below and with solution parameters $\epsilon=0.1$, $ue=-0.1$,
$uc=0.1$. The typical breather oscillation is clearly visible; its period
is of $105.1$ time units ($1 t.u. = 1.02 10^{-14} s$). The amplitude 
$2 \epsilon {\cal A}/ \alpha$ and the half heigth width, which is about 
$2.77 L_e/\epsilon$, are respectively $0.016$ \AA and $20 \,b.p$. }
\end{figure}

\newpage

\begin{figure}[h]
\centerline{\psfig{figure=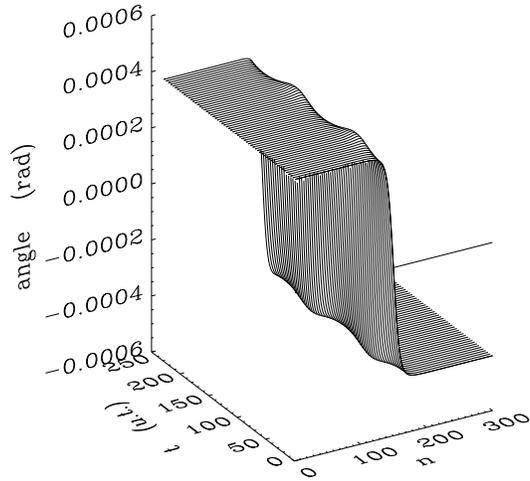,height=8cm}}
\vspace{0.5truecm}
\caption{Fig. 3b: Analytical solution (\ref{y}), (\ref{phi}):  
the corresponding angular 
displacement $\varphi_n(t)$. The main contribution to this solution is
the kink-like term 
$\epsilon \sigma_c L_e {\cal A}^2 \tanh{(\eta(x - V_e t))}$ of  (\ref{phi}).}
\end{figure}

\newpage

\begin{figure}[h]
\centerline{\psfig{figure=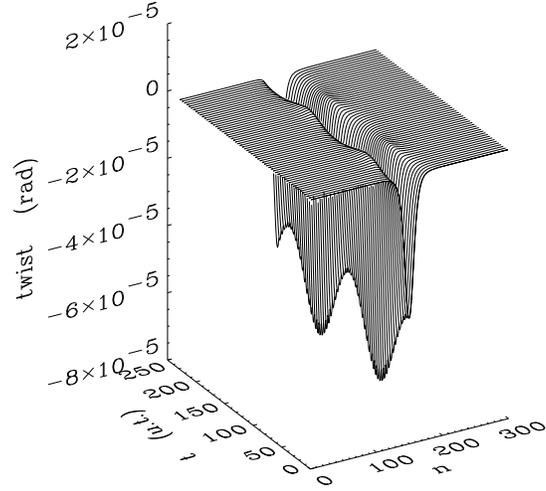,height=8cm}}
\vspace{0.5truecm}
\caption{Fig. 3c: Analytical solution (\ref{y}), (\ref{phi}): 
the twist angle, given by the difference between neighboring  angles,
$\Delta\varphi_n(t)$. We stress 
that the breather opening mode in the radial 
variables corresponds to an untwisted
region.}
\end{figure}

\newpage

\begin{figure}[h]
\centerline{\psfig{figure=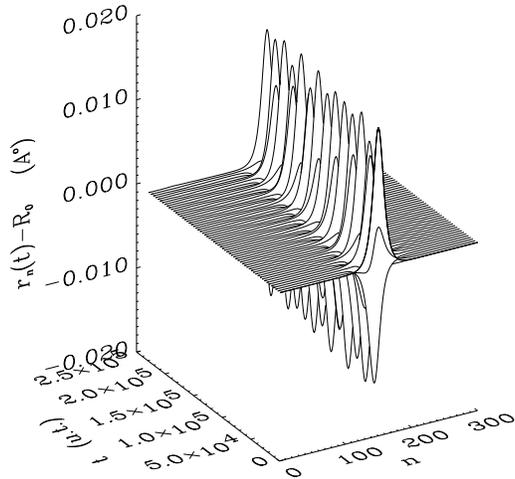,height=8cm}}
\vspace{0.5truecm}
\caption{Fig. 4a: Numerical integration (obtained with a Runge Kutta 
algorithm) of the initial condition which
corresponds to the analytical solution of Fig.~3. The amplitude  and 
width parameters preserve their initial values 
$2 \epsilon {\cal A}  \sim 0.016$ \AA 
and $2.77 L_e/\epsilon \sim  20  \,b.p.$ with good accuracy, as well as
the oscillation period. The total simulation time  corresponds 
to $2234$  oscillation periods. We introduce here absorbing boundary 
conditions to avoid noise effects produced by reflection of 
the small amount of radiated energy 
(less then $ 10^{-5}$ of the total energy).}
\end{figure}

\newpage

\begin{figure}[h]
\centerline{\psfig{figure=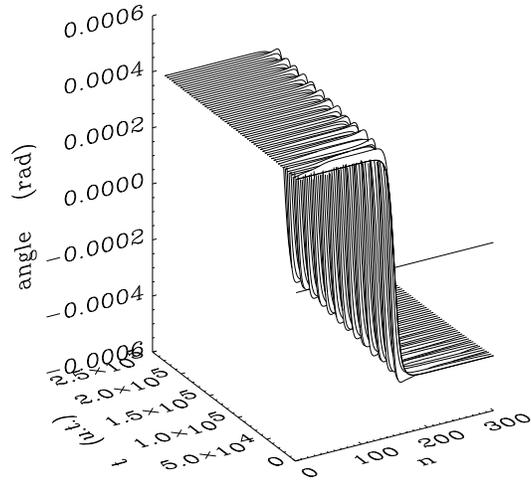,height=8cm}}
\vspace{0.5truecm}
\caption{Fig. 4b:  Angular displacement $\varphi_n(t)$ generated by the
numerical integration of the same initial conditions.}
\end{figure}

\newpage

\begin{figure}[h]
\centerline{\psfig{figure=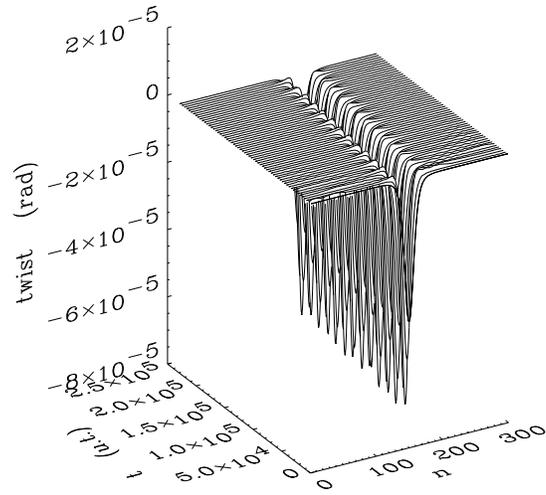,height=8cm}}
\vspace{0.5truecm}
\caption{Fig. 4c: Numerical integration of the same initial conditions, 
 twist angle.}
\end{figure}

\newpage

\begin{figure}[h]
\centerline{\psfig{figure=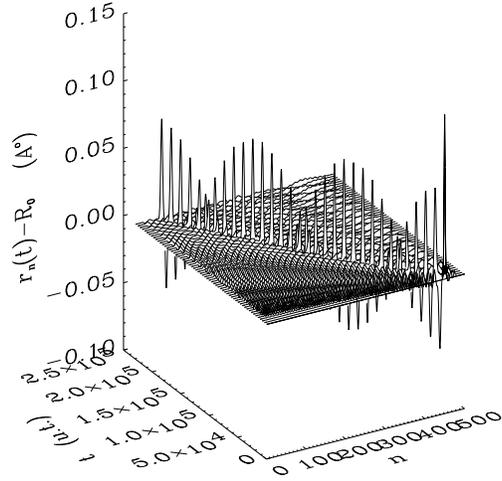,height=8cm}}
\vspace{0.5truecm}
\caption{Fig. 5a: Numerical integration of an initial condition obtained
with solution parameters  $\epsilon=0.2$, $ue=-0.2$,
$uc=0.6$. We perform again the simulation with 
absorbing boundary conditions.  The initial amplitude is 
$2 \epsilon {\cal A} = 0.1$ \AA and the half height width
$2.77 L_e/\epsilon \sim 4\, b.p.$ This initial solution is not stable. 
The initial radiation is clearly visible. However, after some time,
the system tends to stabilize to a smaller breather+kink mode with amplitude
$2 \epsilon {\cal A} \sim 0.07$ and width 
$2.77 L_e/\epsilon \sim 6 \,b.p.$ (We stress that the apparent slow modulation
of the maximum amplitudes and the apparent
oscillation period are due to the beating between the effective 
oscillation and the sampling time.) }
\end{figure}

\newpage

\begin{figure}[h]
\centerline{\psfig{figure=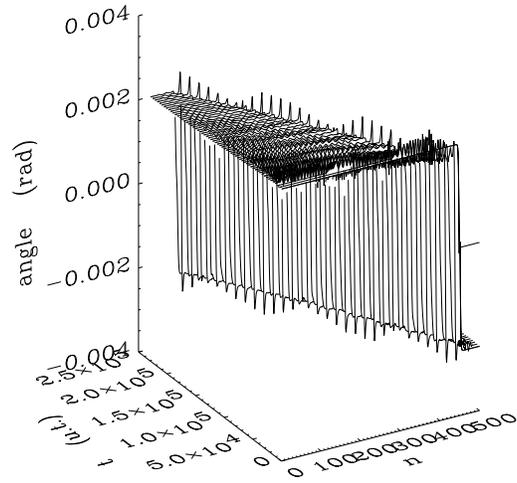,height=8cm}}
\vspace{0.5truecm}
\caption{Fig. 5b: Numerical integration for the same parameters as in 
Fig.~5a, angular displacement $\varphi_n(t)$.}
\end{figure}

\newpage

\begin{figure}[h]
\centerline{\psfig{figure=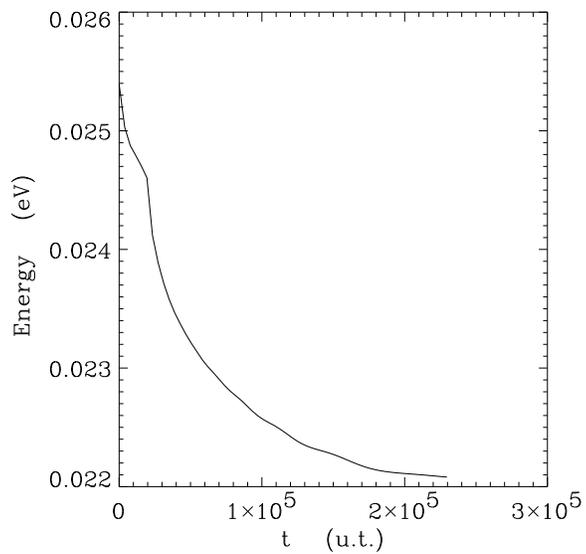,height=8cm}}
\vspace{0.5truecm}
\caption{Fig. 5c: Total system energy  during the simulation with
parameters as in Fig.~5a. The loss of energy is due to the absorbtion 
at the boundaries and represent essentially the radiative part.
 After the initial transition to the smaller 
amplitude breather+kink mode the energy tends to stabilize.}
\end{figure}

\end{document}